\documentclass[9pt,conference]{IEEEtran}
\IEEEoverridecommandlockouts
\usepackage{cite}
\usepackage{amsmath,amssymb,amsfonts}
\usepackage{algorithmic}
\usepackage{graphicx}
\usepackage{textcomp}
\def\BibTeX{{\rm B\kern-.05em{\sc i\kern-.025em b}\kern-.08em
    T\kern-.1667em\lower.7ex\hbox{E}\kern-.125emX}}

\usepackage{adjustbox}

\usepackage{hyperref}
\hypersetup{
    colorlinks=true,
    linkcolor=blue,
    citecolor=red,
    filecolor=magenta,      
    urlcolor=cyan,
    pdftitle={Overleaf Example},
    pdfpagemode=FullScreen,
    }

\usepackage{pifont}% http://ctan.org/pkg/pifont
\newcommand{\cmark}{\textcolor{blue}{\ding{51}}}%
\usepackage{multirow}
\usepackage[symbol]{footmisc}
\usepackage[affil-it]{authblk}
\usepackage{booktabs}
\usepackage[table,xcdraw]{xcolor}

\usepackage{subcaption}

\usepackage{soul}
\usepackage{ifthen}

\usepackage{enumitem}

\usepackage{arydshln}

\usepackage{array}

\definecolor{lightgray}{RGB}{230,230,230}

\newcommand{\thinline}{\arrayrulecolor{black}\specialrule{0.5pt}{0pt}{0pt}\arrayrulecolor{black}}
\newcommand{\thickline}{\arrayrulecolor{black}\specialrule{1.5pt}{0pt}{0pt}\arrayrulecolor{black}}

\newcommand{\midline}{\arrayrulecolor{black}\specialrule{1pt}{0pt}{0pt}\arrayrulecolor{black}}

\usepackage{dsfont}
\begin{document}

\title{Leave-One-EquiVariant: Alleviating invariance-related information loss in contrastive music representations
}

\author[1,2]{Julien Guinot*}
\author[2]{Elio Quinton}
\author[1]{Gy\"orgy Fazekas}

\affil[1]{School of EECS, Queen Mary University of London, London, U.K}
\affil[2]{Music \& Audio Machine Learning Lab, Universal Music Group, London, U.K}

% \and
% \and
% \and
% \and
% \and

\maketitle

\begin{abstract}

Contrastive learning has proven effective in self-supervised musical representation learning, particularly for Music Information Retrieval (MIR) tasks. However, reliance on augmentation chains for contrastive view generation and the resulting learnt invariances pose challenges when different downstream tasks require sensitivity to certain musical attributes. To address this, we propose the Leave One EquiVariant (LOEV) framework, which introduces a flexible, task-adaptive approach compared to previous work by selectively preserving information about specific augmentations, allowing the model to maintain task-relevant equivariances. We demonstrate that LOEV alleviates information loss related to learned invariances, improving performance on augmentation related tasks and retrieval without sacrificing general representation quality. Furthermore, we introduce a variant of LOEV, LOEV++, which builds a disentangled latent space by design in a self-supervised manner, and enables targeted retrieval based on augmentation related attributes.

\end{abstract}


\section{Introduction, Related work}\label{section: Introduction}

Contrastive learning is a powerful paradigm for learning self-supervised representations. These representations have been proven to be effective on a range of downstream tasks, including in MIR. Since its first adaptation from computer vision \cite{chenSimpleFrameworkContrastive2020} by \textit{Spijkervet et al.} \cite{spijkervetContrastiveLearningMusical2021}, contrastive learning of musical representations has been successfully repurposed with different architectures \cite{zhaoS3TSelfSupervisedPretraining2022, vasquezTAILEDUNETMULTISCALE2022}, notably MULE \cite{mccallumSupervisedUnsupervisedLearning2022}), and positive mining strategies \cite{garoufisMultiSourceContrastiveLearning2023,choi2022towards,guinot2024semi,yaoContrastiveLearningPositiveNegative2022, ciranniCOCOLACoherenceOrientedContrastive2024}. Despite its success, the performance of contrastively learned representations on downstream tasks is sensitive to the positive mining strategy and the augmentation chain. Training a contrastive model involves maximizing (resp. minimizing) agreement between positive (resp. negative) augmented samples. Thus, the choice of positive sampling strategy and the augmentations the model learns to ``ignore'' significantly impact the usefulness of learnt representations w.r.t. downstream tasks. For example, in music, genre is transposition invariant: transposing a piece of music does not change its genre. So, including pitch shifting in contrastive training should benefit genre recognition by teaching the model to recognise similar representations across different keys. However, this would harm performance on tasks such as pitch detection, chord recognition or key detection, which rely on key-dependent features. Incorporating an augmentation in training renders representations invariant to this transformation, which may or may not be useful depending on the task. Furthermore, the positive sampling strategy implicitly guides the notion of similarity. That is, a contrastive model trained in this framework is never truly generic, and learnt invariances and similarities might not be task-appropriate for downstream applications.

In practice and to the best of our knowledge, there exists no unified understanding or theory of how the augmentation chain and the positive mining strategy influence downstream performance or each other, nor any one-size-fits-all recipe for contrastive learning approaches. Some works have attempted to alleviate these issues by devising better positive mining strategies \cite{yangClassAwareContrastiveSemiSupervised2022}, reducing false negatives within the mining strategy \cite{huynhBoostingContrastiveSelfSupervised2022,geRobustContrastiveLearning2021}, or influencing positivity and negativity with semantic weighing \cite{guinot2024semi, yangClassAwareContrastiveSemiSupervised2022, wuWav2CLIPLearningRobust2022}. From the standpoint of the augmentation chain, a body of work on understanding the effect of certain augmentations on downstream tasks exists \cite{tianWhatMakesGood2020, xiaoWhatShouldNot2021}, and similar studies have started to appear in MIR \cite{ontheeffect}. Information-theoretical work has attempted to understand the interplay of positive mining and augmentation chains in training contrastive models \cite{tianWhatMakesGood2020}. One work of particular interest to this study, Leave One Out Contrastive (LOOC) \cite{xiaoWhatShouldNot2021}, proposes an approach that does not discard information related to pre-training augmentations in representations. Our work adapts this paradigm to MIR by alleviating learnt invariances pertaining to musical pitch and tempo. Briefly, our contributions are as follows.

% \begin{figure*}
%     \centering
%     \includegraphics[width=.85\linewidth]{pics/MuLOOC.png}
%     \caption{Leave One EquiVariant framework. Each subspace is made invariant to all augmentations but one, forcing the embedding superspace $\mathcal{V}$ to conserve information about all transformations. To do so, augmentations in the variant pipeline are tracked to determine positives and negatives in each projection head}
%     \label{fig:MuLOOC}
% \end{figure*}

\begin{figure}[t]
    \centering
    \includegraphics[width=.8\linewidth]{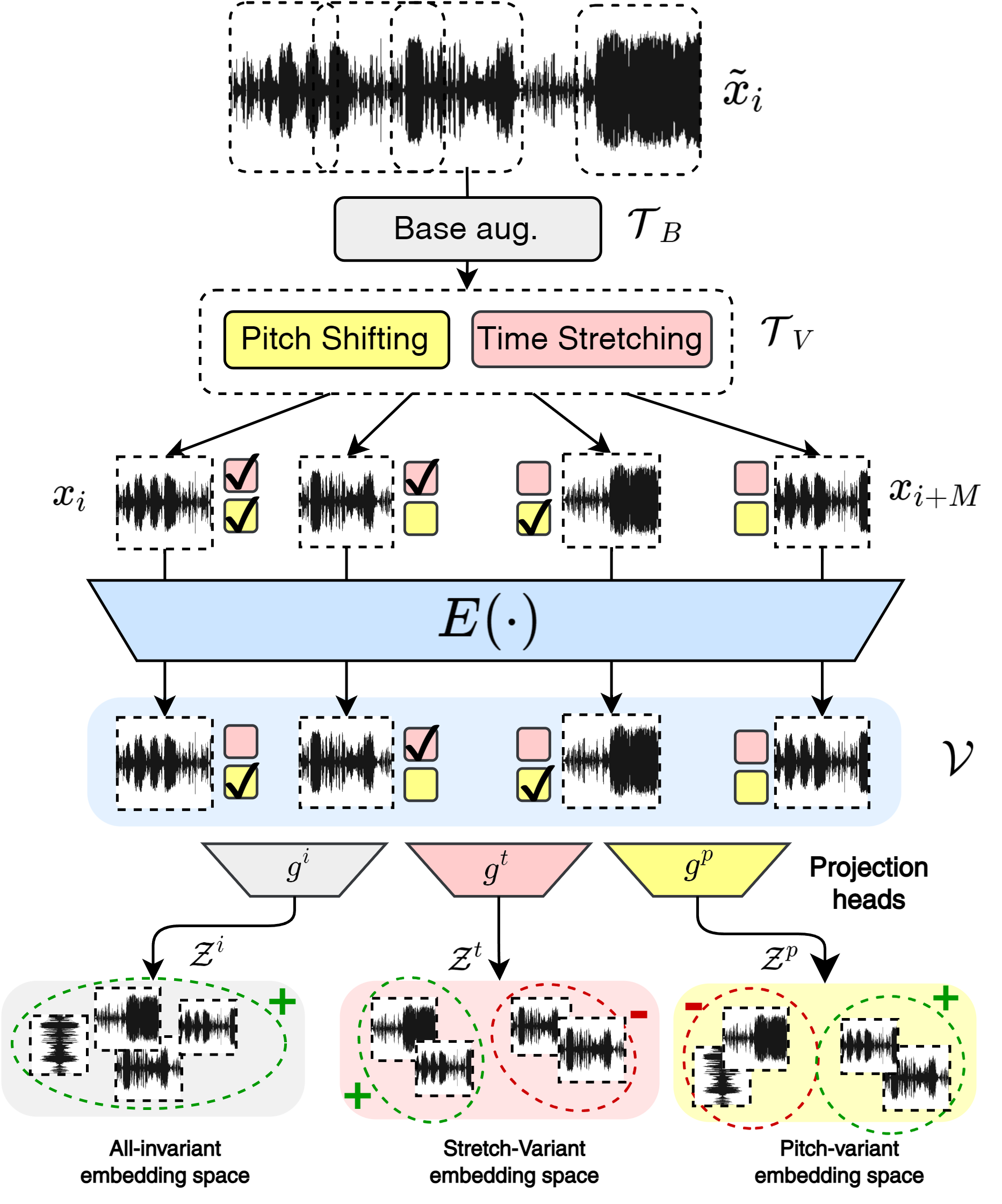}
    \caption{Leave One EquiVariant framework. Subspace $\mathcal{Z}^k$ is invariant to all transformations except $T_k$, forcing the embedding superspace $\mathcal{V}$ to conserve information about all transformations.}
    \vspace{-15pt}
    \label{fig:MuLOOC}
\end{figure}

% \begin{itemize}[left=0pt]
%     \item We propose Leave One EquiVariant (LOEV), and its variant LOEV++, a flexible adaptation of the Leave One Out Contrastive (LOOC) framework, and show its ability to alleviate invariance related information loss with a more flexible ad-hoc approach.
%     \item We show that LOEV(++) improves performance on augmentation-related tasks and retrieval at no cost to general representation potency compared to traditional contrastive approaches. \cite{mccallumSupervisedUnsupervisedLearning2022}.
%     \item We show that LOEV++ builds a disentangled latent space which allows targeting desirable attributes in retrieval and controlling where augmentation information is stored in the latent space.
%     \item Unlike \cite{xiaoWhatShouldNot2021}, where the link between augmentations and fine-grained tasks is unclear, we show that for explicitly linked augmentations and MIR tasks, LOEV(++) significantly reduces performance loss.

% \end{itemize}

\begin{itemize}[left=0pt]
    \item We introduce Leave One EquiVariant (LOEV) and its variant LOEV++, an adaptation of the Leave One Out Contrastive (LOOC) framework, to mitigate invariance-related information loss with greater flexibility.
    \item LOEV(++) enhances performance on augmentation-related tasks and retrieval without compromising general representation quality.
    \item LOEV++ creates a disentangled latent space, enabling targeted retrieval and controlled storage of augmentation information.
    \item Unlike \cite{xiaoWhatShouldNot2021}, where links between augmentations and downstream task are unclear, we demonstrate that LOEV(++) reduces performance loss for explicitly linked augmentations and MIR tasks.
\end{itemize}

\section{Methods}\label{Section: Methods}

% In this work, we leverage an adaptation to the music domain of the leave-one-out contrastive framework proposed in \cite{xiaoWhatShouldNot2021}. Compared to previous work in the image domain where applied augmentations are only loosely linked to important attributes for image classification, we study the case of transformations in the music domain that are directly related to semantic attributes (pitch shifting relating to key and time stretching relating to tempo). The success of MULE \cite{mccallumSupervisedUnsupervisedLearning2022} as a contrastive learning model inspires us to employ the model proposed in \cite{mccallumSupervisedUnsupervisedLearning2022} in this work. 

\subsection{Contrastive embedding space}\label{Subsection: Contrastive embedding space}
 
 In the traditional contrastive learning framework, consider a batch of non-augmented samples $\{\Tilde{x}_i\}, i \in [1...N]$. These samples are augmented through an ordered stochastic chain of $K$ nuclear augmentations $\{T_k\}, k \in [1...K]$. These nuclear augmentations have a probability $p_k$ of being applied. Parameters for each $T_k$ are randomised for each augmented sample. Non-augmented samples $\mathbf{\Tilde{x}}$ are augmented $M$ times into samples $\textbf{x}_i = \{ x_i\}, i \in [1...NM]$. The set of positives $P(i)$ for a given index $i$ is the set of samples originating from the same original sample $x_i$. The contrastive objective is given by the following loss for a sample $i$:

\begin{equation}
    \mathcal{L}_i = \frac{-1}{|P(i)|} \sum_{p \in P(i)} \frac{\exp (sim(z_i,z_p)/ \tau)}{\sum_{j \neq i}\exp (sim(z_i,z_j)/ \tau)}
\end{equation}

$z_i$  is the output of an encoder $E : x \mapsto e \in \mathbb{R}^{d_E}$ and a projection head $g : e \mapsto z \in \mathcal{Z}^i \subset \mathbb{R}^{d_g}$, $\tau$ is a temperature hyperparameter.
% In the case of MULE \cite{mccallumSupervisedUnsupervisedLearning2022}, the encoder is a convolutional F0-NFNet applied to mel-spectrogram features $x \in \mathbb{R}^{96 \times 300}$ (See Section \ref{Subsection: Model input, augmentation chain}). 

In this framework, the network is taught to be invariant to all transformations (all-invariant) by mapping augmented samples to one embedding. We propose a novel, ad-hoc augmentation tracking framework that can be added to existing contrastive learning pipelines where the model maintains information about all augmentations of interest - \textbf{L}eave \textbf{O}ne \textbf{E}qui\textbf{V}ariant (LOEV). In addition to an all-invariant projection head $g^i$ into an embedding space $\mathcal{Z}^i$, $K$ projection heads $g_k$ project the hidden representation $e \in \mathcal{V}$ onto $K$ spaces $\{\mathcal{Z}^k\}$. representations in $\mathcal{Z}^k$ are invariant to all transformations except $T_k$. For each subspace to maintain variance w.r.t a transformation, the set of positives $P_k(i)$ for sample $i$ w.r.t $T_k$ is samples from the same anchor that have \emph{not} been augmented with $T_k$. This way, each projection space is explicitly taught to be variant to \emph{one} augmentation, and the embedding space $\mathcal{V}$ must contain information about all transformations to minimize all objectives. We employ a set of \emph{base} augmentations $\mathcal{T}_B$ without a dedicated subspace (i.e. the model learns to be invariant to them in all subspaces) and \emph{variant} augmentations $\mathcal{T}_V$ which each have a dedicated variant subspace. The contrastive objective for variant subspace $\mathcal{Z}^k$ is:

\begin{equation}
    \mathcal{L}_i^k = \frac{-1}{|P_k(i)|} \sum_{p \in P_k(i)} \frac{\exp (sim(z_i,z_p)/ \tau)}{\sum_{j \neq i}\exp (sim(z_i,z_j)/ \tau)}
\end{equation}

The global loss is then averaged over all subspaces:

\begin{equation}
    \mathcal{L}_i^T = \frac{1}{K+1}  \Bigg( \mathcal{L}_i + \sum_{k=0}^K \mathcal{L}_i^k \Bigg) 
\end{equation}

 \subsection{View generation}\label{Subsection: view generation}
 
 To generate positive views, Leave-one-out contrastive (LOOC) \cite{xiaoWhatShouldNot2021}, keeps the clean anchor as a query; After generating an augmented sample, parameters from each nuclear augmentation except one per view are copied across $K$ additional views. Our framework is similar in spirit but more flexible, employing an ad-hoc positive tracking method as well as a simCLR \cite{chenSimpleFrameworkContrastive2020} contrastive pipeline versus MoCo \cite{chenImprovedBaselinesMomentum2020} in \cite{xiaoWhatShouldNot2021}. Instead of always applying all augmentations but one for $K$ augmentations, which induces some inflexibility to the implementation, we maintain our stochastic augmentation chain and simply track which augmentations are applied to each sample $i$ within a binary vector $t_i \in \{0,1\}^K$ ($t_{i,k} = 1$ if augmentation $k$ is applied to sample $i$ else $0$). For a given augmentation $T_k$, the set of positives is $ P_k(i) = P(i) \cap \{j | t_{j,k} = t_{i,k} = 0 \} $. If two samples have been augmented with the target augmentation, we consider that stochastic uniform sampling of continuous parameters of the augmentation is a sufficient guarantee that the embedding should \emph{not} be the same for both views.

% This is an important design choice, given that this renders our network variant to any parameters of the augmentation while it should also be invariant to the same augmentation applied with the same parameters. The stochastic nature of augmentation can also induce randomly-generated same-parameter same-augmentation views of the same chunk which in our approach would be treated as negatives. While unlikely, these \emph{can} occur and introduce some confusion in the training objective for the model. We might consider an additional augmentation step where, if $t_j^k = t_i^k = 0$, we sample a set of augmentation parameters and apply this same augmentation to both sample $i$ and $j$. Another option is to apply the same augmentation randomly and set $t_j^k = t_i^k = 0$ . Finally, we consider that applying any augmentation to a track should result in an equal repulsion in the latent space, while it may occur that some augmentations are more meaningful than others. For instance, a time stretching factor of 1.01 is perceptually harder to detect than one of 1.2, but in the current framework we consider them to be the same.

\begin{table*}[t]
\centering
\footnotesize
\begin{adjustbox}{max width=0.9\textwidth}
\begin{tabular}{lllccccccccccccc}
\toprule
 & \multicolumn{4}{c}{Aug} & \multicolumn{1}{l}{} & \multicolumn{2}{c}{MTT50} & \multicolumn{2}{c}{Jam50} &  & Giansteps & NSynth &  & \multicolumn{2}{c}{AllTempo} \\ \cline{2-5} \cline{7-10} \cline{12-13} \cline{15-16} 
 & \multicolumn{1}{c}{Mixup} & \multicolumn{1}{c}{B} & PS & TS &  & AUROC & AP & AUROC & AP &  & Acc$_w$ & Acc &  & Acc1 & Acc2 \\ \hline
\multirow{4}{*}{MULE} & \cmark &  &  &  &  & 89.4 \textcolor{red}{($\downarrow$)} & 36.4 \textcolor{red}{($\downarrow$)} &  82.0 & 26.6 &  & 51.9 \textcolor{red}{($\downarrow$)} & 84.0 \textcolor{red}{($\downarrow$)} &  & 69.3 \textcolor{red}{($\downarrow$)} & 90.2 \textcolor{red}{($\downarrow$)} \\
 &  & \cmark &  &  &  & 89.7 \textcolor{blue}{(B)} & 37.3 \textcolor{blue}{(B)} & 82.3 & 27.3 &  & 57.4 \textcolor{blue}{(B)} & 85.4 \textcolor{blue}{(B)} &  & 70.4 \textcolor{blue}{(B)} & 92.3 \textcolor{blue}{(B)} \\
 &  & \cmark & \cmark &  &  & 90.1  \textcolor{green}{($\uparrow$)} & 38.0  \textcolor{green}{($\uparrow$)} &  82.3 & 27.3 &  & 16.7 \textcolor{red}{($\downarrow$)} & 74.8 \textcolor{red}{($\downarrow$)} &  & \textbf{72.1} \textcolor{green}{($\uparrow$)} & \textbf{92.6} \textcolor{green}{($\uparrow$)} \\
 &  & \cmark &  & \cmark &  & 90.3  \textcolor{green}{($\uparrow$)} & 38.4  \textcolor{green}{($\uparrow$)} &  82.4 & 27.3 &  & \textbf{64.7} \textcolor{green}{($\uparrow$)} &  
 \textbf{89.2} \textcolor{green}{($\uparrow$)} &  & 36.0 \textcolor{red}{($\downarrow$)} & 44.7 \textcolor{red}{($\downarrow$)} \\
 &  & \cmark & \cmark & \cmark & & \textbf{90.6} \textcolor{green}{($\uparrow$)} & \textbf{38.7} \textcolor{green}{($\uparrow$)}  & \textbf{82.6} & \textbf{27.7} & & 15.9 \textcolor{red}{($\downarrow$)} &  80.1 \textcolor{red}{$(\downarrow)$} & &  65.1 \textcolor{red}{($\downarrow$)} & 85.3 \textcolor{red}{($\downarrow$)} \\ \hline
MULE++ &  & \cmark & \cmark & \cmark & & 90.5 \textcolor{green}{($\uparrow$)} & 38.4 \textcolor{green}{($\uparrow$)} & 82.7 & 27.8  &  & 15.1 \textcolor{red}{($\downarrow$)} & 78.8 \textcolor{red}{($\downarrow$)} &  & 63.2 \textcolor{red}{($\downarrow$)} & 84.1 \textcolor{red}{($\downarrow$)}   \\ \hline
\multirow{3}{*}{LOEV} &  & \cmark & \cmark &  & & 90.0  \textcolor{green}{($\uparrow$)}  & 38.0  \textcolor{green}{($\uparrow$)}  & 82.4 & 27.3 & & 38.9 \textcolor{red}{($\downarrow$)}  & 84.0 \textcolor{red}{($\downarrow$)}  & &  \textbf{71.3}  \textcolor{green}{($\uparrow$)}  & \textbf{91.9}  \textcolor{red}{($\downarrow$)}\\
 &  & \cmark &  & \cmark &  & 90.0 \textcolor{green}{($\uparrow$)} & 37.8 \textcolor{green}{($\uparrow$)} & 82.3 & 27.2 &  & \textbf{60.0} \textcolor{green}{($\uparrow$)} & \textbf{88.1} \textcolor{green}{($\uparrow$)} &  & 71.2 \textcolor{green}{($\uparrow$)} & 90.8 \textcolor{red}{($\downarrow$)} \\
 &  & \cmark & \cmark & \cmark &  & \textbf{90.5} \textcolor{green}{($\uparrow$)}& \textbf{38.4} \textcolor{green}{($\uparrow$)} & \textbf{82.6} & \textbf{27.7} & & 40.0 \textcolor{red}{($\downarrow$)}  & 83.5 \textcolor{red}{($\downarrow$)} & & 70.0 \textcolor{red}{($\downarrow$)}  & 90.3 \textcolor{red}{($\downarrow$)}\\ \hline
LOEV++ &  & \cmark & \cmark & \cmark &  & 90.6 \textcolor{green}{($\uparrow$)} & 38.4 \textcolor{green}{($\uparrow$)} & 82.7  & 27.8 &  & 44.2 \textcolor{red}{($\downarrow$)}  & 84.5 \textcolor{red}{($\downarrow$)} &  & 72.6 \textcolor{green}{($\uparrow$)} & 91.1 \textcolor{red}{($\downarrow$)} \\ \hline
MULE-L \cite{mccallumSupervisedUnsupervisedLearning2022} & \cmark &  & \multicolumn{1}{l}{} & \multicolumn{1}{l}{} & \multicolumn{1}{l}{} & 91.4 & 40.4 &  &  &  & 66.7 & 89.2 & \multicolumn{1}{l}{} & \multicolumn{1}{l}{-} & \multicolumn{1}{l}{-} \\
MULE-S \cite{mccallumSupervisedUnsupervisedLearning2022} & \cmark &  & \multicolumn{1}{l}{} & \multicolumn{1}{l}{} & \multicolumn{1}{l}{} & 90.5 & 38.9 &  &  &  & 50.8 & 82.4 & \multicolumn{1}{l}{} & \multicolumn{1}{l}{-} & \multicolumn{1}{l}{-} \\
% SOTA & - &  & \multicolumn{1}{l}{} & \multicolumn{1}{l}{} & \multicolumn{1}{l}{} &  &  &  &  &  & 79.6 & 94.4 & \multicolumn{1}{l}{} & \multicolumn{1}{l}{} & \multicolumn{1}{l}{} \\
\bottomrule
\end{tabular}%
\end{adjustbox}
\caption{Results for downstream probing on automatic tagging (MagnaTagATune, Jam50), Key (Giantsteps) and pitch (NSynth) estimation and tempo estimation (AllTempo). Baseline is denoted by \textcolor{blue}{(B)}. Performance improvements (resp. losses) are denoted by \textcolor{green}{($\uparrow$)} (resp. \textcolor{red}{($\downarrow$)}).}
\vspace{-10pt}
\label{tab:results}
\end{table*}

\subsection{Model input, pretraining augmentation chain}\label{Subsection: Model input, augmentation chain}

Our model (Section \ref{subsection: Model Architecture}) takes as inputs log-scaled mel-spectrograms. We sample 3 seconds of 16kHz mono audio and convert it to a Log-Mel Spectrogram. Prior to the augmentation chain, $M$ chunks of 3 seconds are sampled from the track according to 3 strategies and then augmented: ``\emph{Same}'' strategy samples the same chunk $M$ times. ``\emph{Adjacent}' strategy samples $M$ adjacent chunks with no overlap. ``\emph{Random}'' strategy randomly samples $M$ chunks from the track, allowing for overlap. These strategies accommodate previous findings that different positive positions in the track lead to different representation strengths \cite{choi2022towards}. As key and tempo are \emph{relatively} position-invariant within a track, we assume this will not have an impact on learned representations. $\mathcal{T}_B$ includes Gain, Polarity Inversion, Colored Noise addition, Filtering (one of low / high passing, or band cut / passing), Reverb, and Distortion. $\mathcal{T}_V$ includes Pitch shifting continuously between -4 and 4 semitones, and Continuous time stretching with factors sampled between 0.7 and 1.3 -  Both applied with 50\% chance. We augment each anchor sample 4 times. 

\subsection{Model architecture}\label{subsection: Model Architecture}

As in \cite{wangLearningUniversalAudio2022,mccallumSupervisedUnsupervisedLearning2022}, we leverage a F0-SF-NFNet as the encoder for our contrastive task.
Official implementations are available only in Jax \footnote{\hyperlink{https://github.com/google-deepmind/slowfast_nfnets}{https://github.com/google-deepmind/slowfast\_nfnets}} and Keras \footnote{\hyperlink{https://github.com/PandoraMedia/music-audio-representations}{https://github.com/PandoraMedia/music-audio-representations}}, so we reproduce the model in Pytorch.
% Despite our best efforts at reproducing the architecture, we find a 10M parameter discrepancy between our implementation and claimed parameter counts in the literature (63M parameters in \cite{mccallumSupervisedUnsupervisedLearning2022,wangLearningUniversalAudio2022} and 53M in ours).
In MULE \cite{mccallumSupervisedUnsupervisedLearning2022}, the projection head is a 47M parameter 3-layer MLP. Because we use multiple projection heads, we scale back our projection head sizes to two 2048-wide hidden layers. MULE benefits from large-scale high-quality pretraining data, which allows it to forgo traditional augmentation approaches and limit itself to an efficient in-batch mixup approach. We \emph{do} use other augmentations (Section \ref{Subsection: Model input, augmentation chain}).
% Further constraints such as compute budget limit us in terms of batch size, a crucial component to contrastive learning. The differences with the original implementation are summarized in Table \ref{tab:diff}.
% Despite these differences, we manage to reach comparable performance to MULE on a range of downstream tasks mentioned in the paper as shown in Table \ref{tab:results}. Because we are not attempting to surpass the official implementation of MULE but rather to demonstrate the capacity of our method when compared to similarly scaled and trained models, this similar performance is satisfactory for the sake of our study. 

As in \cite{xiaoWhatShouldNot2021}, we implement variations of MULE and LOEV, namely MULE++ and LOEV++. Authors in \cite{xiaoWhatShouldNot2021} introduce LOOC++, which integrates the last convolutional block of the ResNet50 backbone into the projection heads, creating multiple $\mathcal{V}$ superspaces and disentangling in the global latent space into smaller transformation-variant spaces. We do the same for LOEV++ by resorbing a 10M parameter portion of the network into the projection heads. For the all-invariant head, the pitch variant head and the stretch-variant head (resp. $g^i, g^p, g^t$) the spaces prior to the resorbed portion of the model  are notated $\mathcal{V}^i,\mathcal{V}^t,\mathcal{V}^p$, and can all be used as frozen representation spaces (See Sec. \ref{subsection: representation embedding space for downstream probing}). The concatenation of these spaces is $\mathcal{V}^{++}$
% We show the modified LOEV++ architecture in Fig. \ref{fig:MuLOOC++}

% \begin{figure}
%     \centering
%     \includegraphics[width=1\linewidth]{pics/MuLOOC++.png}
%     \caption{LOEV(++) architectures. In either case, the probing representation can come from the embedding superspace, or, for LOEV++, the concatenation of the parallel superspaces can also be used, as in \cite{xiaoWhatShouldNot2021}}
%     \label{fig:MuLOOC++}
% \end{figure}

\section{Experiments and results}\label{Section:Experiments and results}

We study two variant augmentations, pitch shifting (PS) and time stretching (TS). Both these augmentations are single-parameter transformations which are directly correlated to semantic musical information (pitch and key, tempo). These choices illustrate the effectiveness of our method for explicitly semantic audio transformations, but the method itself is transformation-agnostic. 

\subsection{Datasets and evaluation metrics}\label{subsection: Datasets}
\label{sec:datasets}

We use the \emph{MTG-Jamendo dataset} \cite{bogdanov2019mtg} for pretraining. We evaluate pretrained models on automatic tagging, our proxy task for general music understanding, by predicting the top 50 tags for Jamendo and \emph{MagnaTagATune} \cite{MTAT} with canonical data splits \cite{pons2017end,spijkervetContrastiveLearningMusical2021}. We report AUROC and mean Average Precision. Two straightforward transposition-variant tasks are key estimation and pitch estimation. Key estimation is formulated as a 24-way classification task with the \emph{Giantsteps} dataset \cite{knees2015two} for training and \emph{MTG-Giantsteps} \cite{korzeniowski2017end} for testing, as in \cite{yuanMARBLEMusicAudio2023}. The metric for key estimation is a weighted accuracy taking into account reasonable errors \cite{raffel2014mir_eval}. We use the \textit{mir-eval} implementation. We employ \emph{NSynth} \cite{engel2017neural} for pitch estimation, the metric is accuracy over 112 pitch classes.  Tempo estimation is our stretch-variant task. Four datasets : \emph{GTZAN} \cite{tzanetakis2002musical}, \emph{ACM-MIRUM} \cite{peeters2012perceptual}, \emph{Hainsworth} \cite{hainsworth2004particle} and \emph{Giantsteps}  \cite{knees2015two} are used in a one-vs-all fashion, i.e., when testing on one dataset, we train on all 3 other datasets --- we call this dataset \emph{AllTempo} as in \cite{mccallum2024similar}. During probe training, We implement a time-stretching augmentation with a stretching rate $\tau \sim \mathcal{U}(0.8,1.2)$ for robustness \cite{mccallum2024similar,quintonEquivariantSelfSupervisionMusical2022}.  The metrics for this task are 300-class $acc_1$ and $acc_2$, two tolerance-reinforced accuracy metrics \cite{raffel2014mir_eval}. $acc_2$ allows for octave errors, i.e. predictions within a tolerance interval around reasonable ratios of the ground truth.
% Table \ref{tab:datasets} summarizes the datasets leveraged in this work.
\subsection{Pretraining details}\label{subsection: pretraining details}

We pretrain our models on the MTG-Jamendo dataset for 200000 steps with the Adam optimiser. The authors of \cite{mccallumSupervisedUnsupervisedLearning2022} use a cosine decay learning rate scheduler with linear warmup. We find that in our case, perhaps due to the reduced batch size, such scheduling leads to slightly worse results across the board, so we train with a constant learning rate of 0.0001. The authors in \cite{mccallumSupervisedUnsupervisedLearning2022} propose two MULE models trained on proprietary datasets and batch sizes of different scales. We pretrain on comparatively smaller batch sizes (256 vs MULE-Large's 3840 and MULE-Small's 512) and data scale (2.9k hours vs 5k --- MULE-S and 117.5k --- MULE-L ). This partly explains the discrepancies between our results and those reported in \cite{mccallumSupervisedUnsupervisedLearning2022}. Table \ref{tab:results} shows that including time stretching and pitch shifting in the pretraining augmentations compensates for the performance loss due to this down-scaling and is thus an attractive strategy.
% We report differences in training protocols, training scales, and model architectures in Table \ref{tab:diff} to explain possible discrepancies between our results and the results reported in \cite{mccallumSupervisedUnsupervisedLearning2022}.

% % Please add the following required packages to your document preamble:
% % \usepackage{graphicx}
% \begin{table}[h]
% \resizebox{\columnwidth}{!}{%
% \begin{tabular}{llccc}
% \hline
%  &  & \textbf{MULE-Large} & \textbf{MULE-Small} & \textbf{Ours} \\ \hline
% Architecture & #Params & 63M & 63M & 53M \\
%  & Proj. head & \multicolumn{2}{c}{[4096,4096,4096,1728]} & [1728,1728,512] \\ \hline
% Data & Hours & 117.5k & 5k & 2.9k \\
%  & Samples & - & 1.8M & 55k \\ \hline
% Hyperparams & batch size & 3840 & 512 & 256 \\
%  & GPUs & 16 A100 80GB & 16 A100 80GB & 2 A5000 24GB \\
%  & Precision & - & - & Mixed 16bit \\ \hline
% % Aug. & Mixup & \cmark & \cmark & \xmark \\
% %  & Ours - base & \xmark & \xmark & \cmark \\
% %  & Pitch shifting & \xmark & \xmark & \cmark \\
% %  & Time Stretching & \xmark & \xmark & \cmark \\ \hline
% \end{tabular}%
% }}
% \caption{Training and architecture differences between official MULE implementation \cite{mccallumSupervisedUnsupervisedLearning2022} and ours.}
% \label{tab:diff}
% \end{table}

% \subsubsection{Training protocol discussion}
% \label{subsubsection: training protocol discussion}

We pretrain the following model-augmentation chain combinations and use their frozen representations from $\mathcal{V}$ or $\mathcal{V^{++}}$ in Sections \ref{subsection: downstream evaluation},\ref{subsection: representation embedding space for downstream probing} and \ref{subsection: retrieval experiments}: A MULE model trained with only Mixup as the augmentation pipeline, as in the original implementation \cite{mccallumSupervisedUnsupervisedLearning2022}. The parameters for the mixup gain are taken from \cite{mccallumSupervisedUnsupervisedLearning2022}. MULE and LOEV models trained with the augmentation chain specified in Section \ref{Subsection: Model input, augmentation chain}. Used pretraining augmentation chains include \emph{without} variant augmentations (B), \emph{with either} variant augmentations (PS or TS), and \emph{with both} variant augmentations (PSTS). When only one variant augmentation is present, the projection head for the missing variant augmentation is not considered in the loss computation. Finally, we pretrain LOEV++ and MULE++ models with PSTS.

\subsection{Downstream probing evaluation}\label{subsection: downstream evaluation}

For our first set of experiments, we probe frozen representations on tasks pertaining to pitch shifting and time stretching: Tempo estimation, key and pitch estimation, and automatic tagging (See Sec. \ref{sec:datasets}). We train shallow Relu-nonlinear MLPs with different layer depths, widths, and dropout values on frozen representations for each task. We employ the Adam optimizer to train the probes. Dropout, learning rate, and probe architecture are empirically adjusted for each dataset to avoid overfitting. We train on chunk-level embeddings, and average embeddings across the full track for test-time inference, as is customary. Results are reported in Table \ref{tab:results}.

\begin{figure*}[t]
    \centering
    \includegraphics[width=.27\textwidth]{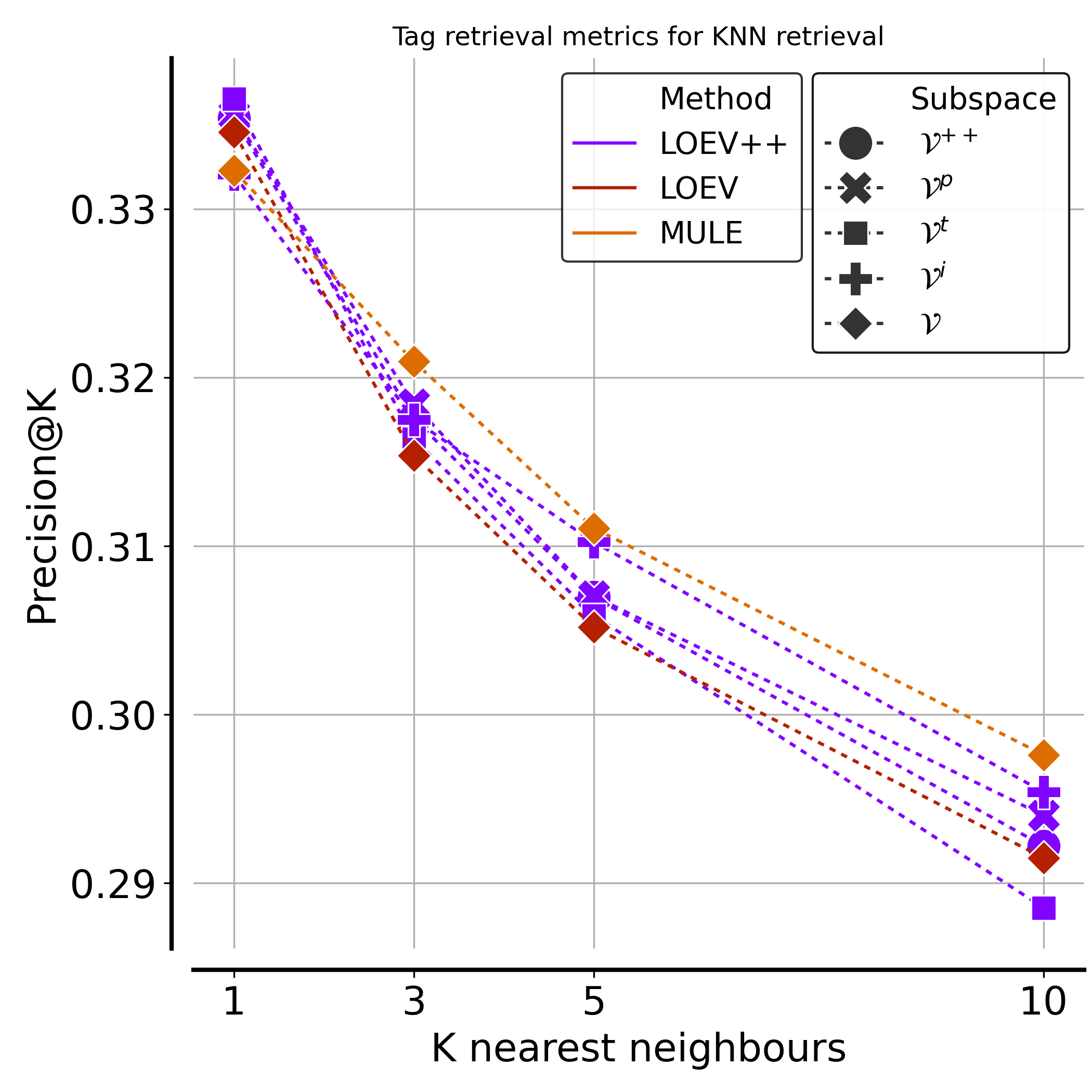}
    \hfill
    \includegraphics[width=.27\textwidth]{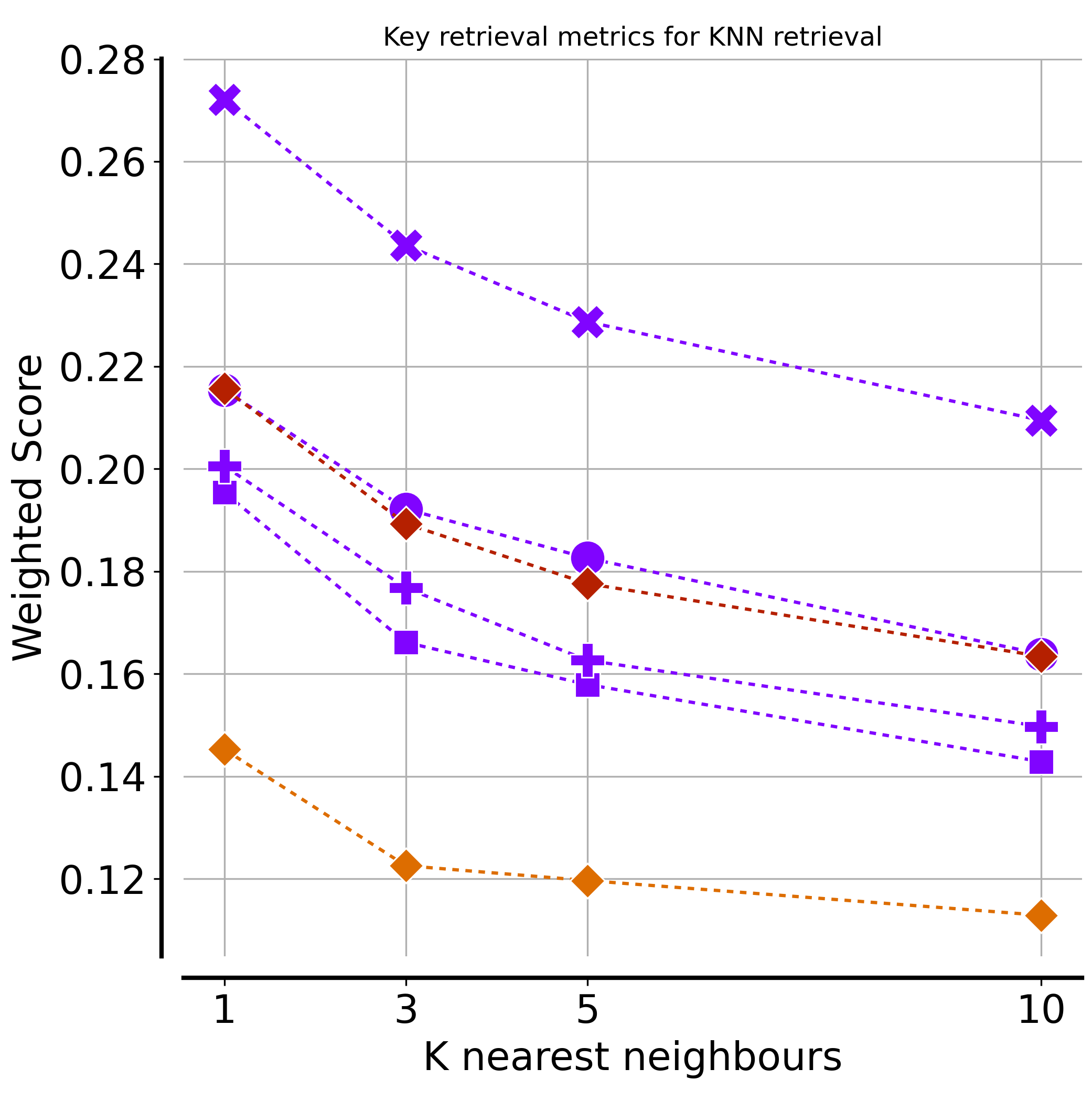}
    \hfill
    \includegraphics[width=.27\textwidth]{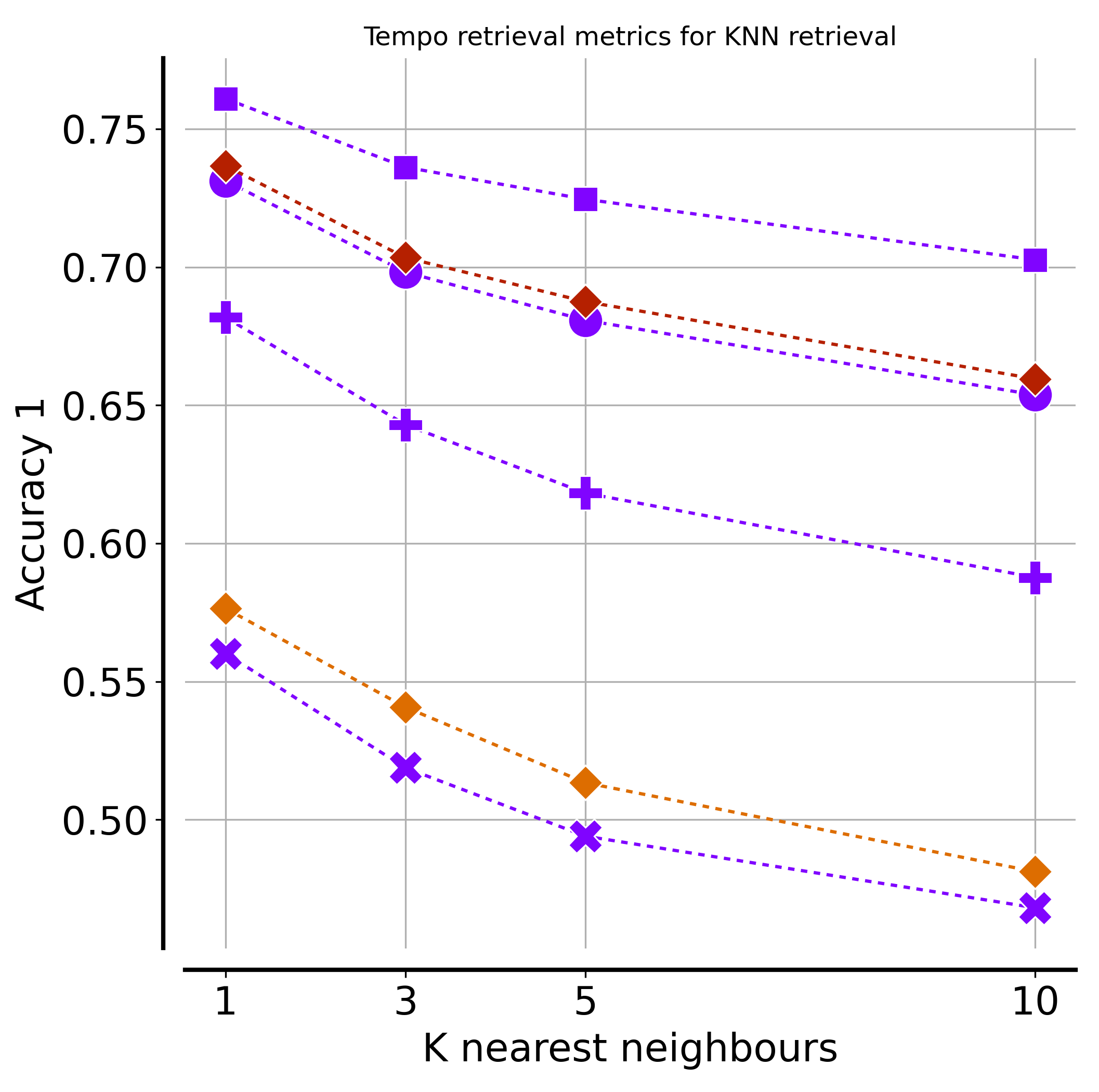}
    \caption{Retrieval metrics for retrieved tags (MagnaTagATune), key (Giantsteps), and tempo (AllTempo) - Precision@K, Weighted accuracy and $acc_1$ are computed between the seed track embedding and the retrieved labels for the $k \in [1,3,5,10]$ nearest neighbouring embeddings.}
    \label{fig:retrieval}
    % \vspace{-3pt}
\end{figure*}

As in previous studies \cite{xiaoWhatShouldNot2021, ontheeffect}, including augmentations in the augmentation chain for the standard contrastive learning method is harmful to tasks that require that information, e.g. pitch shifting with key estimation and time stretching with tempo estimation. (See Table \ref{tab:results})  In counterpart, including these augmentations is beneficial to tasks such as automatic tagging that are key and pitch-invariant to an extent. Note that MULE-Mixup results are always slightly worse than results for MULE-B because our base augmentation chain is \emph{non-destructive} to key and tempo, whereas mixup \emph{is} . The tradeoff between general and task-specific performance is alleviated by LOEV: even with harmful augmentations, the performance drop is drastically reduced. leaving the transformation out entirely still leads to better performance, arguably due to design considerations and hyperparameters for our positive selection strategy. One anomaly is the performance of MULE-TSPS on tempo estimation, which surprisingly does not suffer as much as with MULE-TS. Arguably, a frequency-wise shift is an easier identifiable transformation than a time-wise stretch on 3-second snippets, which the model might be overfitting to. This warrants further investigation.

\subsection{Representation Embedding space for downstream probing}\label{subsection: representation embedding space for downstream probing}

% \textcolor{red}{[In short : try concatenation of different embedding spaces (evaluating performance with all concatenated variations and MULE++, variance/cosine similarity for different augmentations. this checks for disentanglement). Also, look at the activations for different datasets on different subspaces (typically tempo might help a bit on genre, pitch might help more on vocalist or instrument classification)]}

We evaluate the performance of LOEV++ (PSTS) when compared to MULE++ (PSTS) on the same downstream tasks as previously, when probing different subspaces. We report performance on MTAT, average performance over AllTempo and key estimation performance. Results are reported Table \ref{tab:subspace}.  Furthermore, we report performance for LOEV++ when probing $\mathcal{V}^{i}$, $\mathcal{V}^p$, $\mathcal{V}^t$. MULE++ is trained with the same head duplication scheme as LOEV++ but the same objective for all heads, as in \cite{xiaoWhatShouldNot2021}. With this, the probing spaces for MULE++ and LOEV++ are equally sized for fair comparison.

% \usepackage{multirow}
% \usepackage{graphicx}
\begin{table}[h]
\resizebox{\linewidth}{!}{%
\begin{tabular}{lllccccccccc}
\toprule
 &  &  & \multicolumn{2}{c}{MTAT} &  & Giantsteps &  & \multicolumn{2}{c}{AllTempo} \\ \cline{1-2} \cline{4-5} \cline{7-7} \cline{9-11} 
Model & Subspace &  & AUROC & AP &  & Acc$_w$ &  & Acc1 & Acc2 \\ \cline{1-2} \cline{4-5} \cline{7-7} \cline{9-11} 
% MULE & $\mathcal{V}$ &  & 90.6   & 38.7  &  & 15.9 &  & 65.1 & 85.3  \\
% LOEV & $\mathcal{V}$ &   & 90.5 & 38.4 &  & 40.0 &  & 70.0 & 90.3 \\ \hline
MULE++ & $\mathcal{V^{++}}$ &    & 90.5 & 38.4 &  & 15.1 &  & 63.2 & 84.1 \\ \hline
\multirow{7}{*}{LOEV++} & $\mathcal{V^{++}}$ &  & \textbf{90.6}   & \textbf{38.4}  &  & \textbf{44.2} &  & \textbf{72.6} & 91.1 \\
 & $\mathcal{V}^i$ && \textbf{90.5} & \textbf{38.4} &  & 39.0 &    & 70.7 & 89.4   \\
 & $\mathcal{V}^p$ && 90.4 & 38.3 &  & \textbf{43.0} &    & 64.7 & 84.3  \\
 & $\mathcal{V}^t$ && 90.3 & 38.2 &  & 30.0 &    & \textbf{71.5} & \textbf{91.2}  \\
 % & $\mathcal{V}^i \cup \mathcal{V}^p$ &    &  &  &  &  &  &  &  \\
 % & $\mathcal{V}^i \cup \mathcal{V}^t$ &    &  &  &  &  &  &  &  \\
 % & $\mathcal{V}^p \cup \mathcal{V}^t$ &   &  &  &  &  &  &  &  \\
 \bottomrule 
 
\end{tabular}%
}

\caption{Probing subspace experiment results. we compare probing results on automatic tagging, key estimation and tempo estimation while probing different subspaces of $\mathcal{V}^{++}$, namely $\mathcal{V}^i$,$\mathcal{V}^p$,$\mathcal{V}^t$}
\vspace{-10pt}

\label{tab:subspace}
\end{table}

\begin{figure}
    \centering
    \includegraphics[width=1\linewidth]{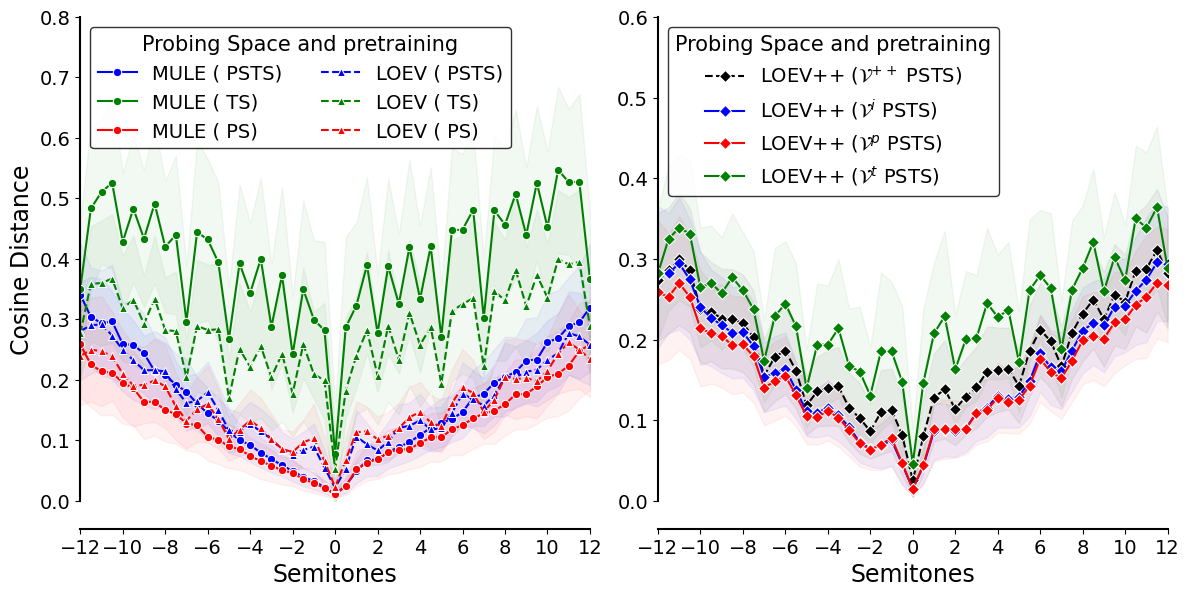}
    \caption{Cosine distance between embeddings of pitch-shifted and non-pitched audio snippets for LOEV and MULE (left) and LOEV++ (right) and different subspaces and pretraining.}
    \label{fig:cosine distance - MULE LOEV}
    \vspace{-10pt}
\end{figure}

We also aim to understand how the information relating to different transformations is stored within the embeddings. We iterate over the MTAT test set and apply both time stretching and pitch shifting. We average the non-augmented and the augmented chunk embeddings $z$ and $z^\star$ over tracks, for different subspaces. We compute the cosine distance $d_c$ between clean and transformed embeddings: $d_c(z,z^{\star}) =1 -  (z \cdot z^\star) / (||z|| ||z^\star||)$ as in \cite{ontheeffect}. The evolution of this cosine distance with the number of semitones applied for pitch shifting is shown in Fig. \ref{fig:cosine distance - MULE LOEV} for MULE and LOEV with various pretraining augmentations and for different subspaces of LOEV++\footnote{We choose to represent pitch shifting as it is more visually explicit, but perform the same experiments for time stretching and find similar results.}. 

% And the standard deviation of each embedding dimension resulting from the difference of the transformed and non-transformed embeddings. This allows us to test for the locality of the information relating to each transformation within the latent space.\\

% \dummyfigure[0.75]{Cosine similarity between embeddings of transformed and non-transformed audio snippets for different models and subspaces}{Line plot}

% \begin{figure}[t]
%     \centering
%     \includegraphics[width=.85\linewidth]{pics/cosine distance.png}
%     \includegraphics[width=.85\linewidth]{pics/pluspluscosine.png}
%     \caption{Cosine similarity between embeddings of transformed and non-transformed audio snippets for MULE, LOEV, and LOEV++ - the applied effect is pitch shifting}
%     \label{fig:cosine distance - MULE LOEV}
% \end{figure}

% \begin{figure}
%     \centering
    
%     \caption{Cosine similarity between embeddings of transformed and non-transformed audio snippets for different probing spaces of LOEV++ - The applied effect is pitch shifting.}
%     \label{fig:cosine distance - LOEV++}
% \end{figure}

Similar to results presented in \cite{ontheeffect}, we find that both MULE-TS and LOEV-TS present more marked fluctuation of cosine distance when pitch shifting the audio than pretrained with PS or PSTS, due to learned invariances. We notice that $d_c(z,z^{\star})$ is higher for MULE than for LOEV, while the general structure is conserved, explaining differences in performance in Table \ref{tab:results}. We verify musically plausible results shown in \cite{ontheeffect} showing noticeable dips in cosine distance at harmonic series pitch shifting factors
% ($\pm 2$ semitones, major second. $\pm5$ semitones, perfect 4th. $\pm7$ semitones, perfect fifth. $\pm12$ semitones, octave)
. In the latent space of LOEV++, we find a similar pitch-variant structure in the $\mathcal{V}^p$ subspace despite training with PSTS, showing that pitch-variant information is well maintained in this space, and confirming findings in Table \ref{tab:subspace}. $d_c(z,z^{\star})$ in other subspaces is smoother and resembles $d_c(z,z^{\star})$ found in $\mathcal{V}$ for LOEV and MULE PS or PSTS.

% \dummyfigure[0.75]{Standard deviation of difference between embedding dimensions of augmented vs non-augmented embeddings for different models and probing spaces.}{Bar plot?}\\

% In addition, we run a simple linear probing experiment on the same datasets as previously and record the average activation values over the different subspaces for different models. This should show that information relating to different semantic concepts relating to the audio is localized depending on the training method.

% \dummyfigure[0.9]{Average activation over different subspaces for different datasets, training methods and subspaces.}{Bar chart}

\subsection{Retrieval experiments}\label{subsection: retrieval experiments}

% \textcolor{red}{[In short : a mix of what is done in \cite{mccallum2024similar} and \cite{xiaoWhatShouldNot2021}]. Essentially, retrieve nearest neighbours and show more consistency along key and tempo dimensions compared to previous baselines.}

To verify that retrieval in the latent space preserves semantic information relating to the applied variant transformations, we run retrieval experiments on our key, tempo, and tagging datasets. We retrieve test set nearest neighbours using cosine distance (KNN with $k \in [1,3,5,10]$) and compute relevant metrics (Table \ref{subsection: Datasets}) between query and retrieved data points averaged over queries. For tempo datasets, We perform KNN retrieval on all 4 datasets simultaneously. We also retrieve KNN tags from the MTAT test set and compute precision @k for each tag, averaged over tags. The intuition behind these experiments is that nearest neighbours in a space which discards information related to a given semantic aspect will not be neighbours based on that information. These experiments are of interest as one of our focuses with this study is manipulating semantic retrieval in the latent space. Fig. \ref{fig:retrieval} reports results for tempo, key, and tag retrieval.

We find that while MULE-PSTS maintains an edge on tag retrieval, it is largely outperformed by both LOEV and LOEV++ subspaces for tempo and key retrieval.
% Further, we find that LOEV(++) suffers from less performance degradation when dealing with the embeddings of transformed audio compared to MULE when retrieving tracks according to key and tempo.
Finally, We find that LOEV++ is also effective in disentangling information in different subspaces for retrieval. Key retrieval yields the best result in the pitch-variant space $\mathcal{V}^p$, and similarly $\mathcal{V}^t$ performs the best for tempo retrieval.

\section{Conclusion and future work}

This paper introduced LOEV(++), a novel method for contrastive representation learning, applied to contrastive learning of musical representations. By incorporating adaptive ad-hoc augmentation tracking and specific augmentation-variant subspaces and training objectives during pretraining, LOEV(++) effectively reduces information loss related to learned invariances in both shallow probing and retrieval tasks with no detriment to tagging performance. LOEV(++) offers performance gains over MULE-PSTS while remaining computationally sober compared to approaching similar performance With MULE-Mixup. We show that LOEV++ explicitly disentangles augmentation information in the latent space, enabling attribute-targeted retrieval and competitive performance on downstream tasks.

This study primarily focused on audio transformations related to key, pitch, and tempo, chosen for their direct links to semantic information. While ideal for demonstrating our method's effectiveness, other transformations such as distortion, reverb, and compression, as well as nonparametric semantic transformations relating to genre or instrumentation, were not explored and are left to future work.

\clearpage

% \bibliographystyle{IEEEbib}
% \bibliography{refs}

% \clearpage

% \section*{Acknowledgment}

% The preferred spelling of the word ``acknowledgment'' in America is without 
% an ``e'' after the ``g''. Avoid the stilted expression ``one of us (R. B. 
% G.) thanks $\ldots$''. Instead, try ``R. B. G. thanks$\ldots$''. Put sponsor 
% acknowledgments in the unnumbered footnote on the first page.

% \section*{References}

% Please number citations consecutively within brackets \cite{b1}. The 
% sentence punctuation follows the bracket \cite{b2}. Refer simply to the reference 
% number, as in \cite{b3}---do not use ``Ref. \cite{b3}'' or ``reference \cite{b3}'' except at 
% the beginning of a sentence: ``Reference \cite{b3} was the first $\ldots$''

% Number footnotes separately in superscripts. Place the actual footnote at 
% the bottom of the column in which it was cited. Do not put footnotes in the 
% abstract or reference list. Use letters for table footnotes.

% Unless there are six authors or more give all authors' names; do not use 
% ``et al.''. Papers that have not been published, even if they have been 
% submitted for publication, should be cited as ``unpublished'' \cite{b4}. Papers 
% that have been accepted for publication should be cited as ``in press'' \cite{b5}. 
% Capitalize only the first word in a paper title, except for proper nouns and 
% element symbols.

% For papers published in translation journals, please give the English 
% citation first, followed by the original foreign-language citation \cite{b6}.

\bibliographystyle{IEEEbib}
\bibliography{refs}
% \end{thebibliography}

\clearpage

\twocolumn[
\hrule
    \begin{center}
        {\Large \bfseries Appendix for the paper ``Leave-One-EquiVariant: Alleviating invariance-related information loss in contrastive music representations'' --- preprint only}
    \end{center}
    \hrule
    \bigskip
    \bigskip
]

\section{Appendix}

\subsection{Training protocol comparison with original MULE}

We report differences in training protocols, training scales, and model architectures in Tab. \ref{tab:diff} to explain possible discrepancies between our results and the results reported in \cite{mccallumSupervisedUnsupervisedLearning2022}.

% \usepackage{graphicx}
\begin{table}[h]
\resizebox{\columnwidth}{!}{
\begin{tabular}{llccc}
\hline
 &  & \textbf{MULE-Large} & \textbf{MULE-Small} & \textbf{Ours} \\ \hline
Architecture & \#Params & 63M & 63M & 53M \\
 & Proj. head & \multicolumn{2}{c}{[4096,4096,4096,1728]} & [1728,1728,512] \\ \hline
Data & Hours & 117.5k & 5k & 2.9k \\
 & Samples & - & 1.8M & 55k \\ \hline
Hyperparams & batch size & 3840 & 512 & 256 \\
 & GPUs & 16 A100 80GB & 16 A100 80GB & 2 A5000 24GB \\
 & Precision & - & - & Mixed 16bit \\ \hline
% Aug. & Mixup & \cmark & \cmark & \xmark \\
%  & Ours - base & \xmark & \xmark & \cmark \\
%  & Pitch shifting & \xmark & \xmark & \cmark \\
%  & Time Stretching & \xmark & \xmark & \cmark \\ \hline
\end{tabular}%
}
\caption{Training and architecture differences between official MULE implementation \cite{mccallumSupervisedUnsupervisedLearning2022} and ours.}
\label{tab:diff}
\end{table}

Despite these differences, we manage to reach comparable performance to MULE on a range of downstream tasks mentioned in the paper as shown in Tab. \ref{tab:results} when training with PSTS, PS or TS. Because we are not attempting to surpass the official implementation of MULE but rather to demonstrate the capability of our method when compared to similarly scaled and trained models, this similar performance serves more as a sanity check and is satisfactory within the scope of this study.

\subsection{Specifics of augmentation chains for pretraining}

Tab. \ref{Tab: augmentations} reports the base augmentation chain $\mathcal{T}_B$ and variant augmentation chain $\mathcal{T}_V$ as well as maximum and minimum transformation parameters:

\begin{table}[h]
\centering
\resizebox{\linewidth}{!}{%
\begin{tabular}{lllll}
\textbf{Augmentation} & \textbf{probability} & \textbf{parameter} & \textbf{Min/Max} & \textbf{unit} \\ \thickline
\\
$\mathcal{T}_B$&&&& \\ \midline

Gain & 0.7 & Gain & -15 / 5  & dB \\ \thinline
Polarity inv. & 0.8 & - & - & - \\ \thinline
Colored Noise & 0.8 & Signal/noise ratio & 3 / 30  & dB \\
 &  & Spectral decay & -2 
/ 2 & dB/octave \\ \thinline
\textit{Filtering} & (One of)  &  &  &  \\ 
Low pass & 0.5 & Cutoff & 0.15 / 7 & kHz \\
High pass & 0.5 & Cutoff & 0.2 / 2.4 & kHz \\
Band pass & 0.5 & center frequency & 0.2 / 4  & kHz \\
 &  &  Bandwidth fraction & 0.5 / 2 & - \\
Band cut & 0.3 & center frequency & 0.2 / 4 & kHz \\
 &  &  Bandwidth fraction & 0.5 / 2 & - \\ \thinline
Reverb & 0.5 & room size & 0.2 / 1 & - \\ 
& & wet factor & 0 / 1  & ratio \\
\thinline
Distortion & 0.6 & Drive & 1 / 10 & dB \\
\\

$\mathcal{T}_V$&&&& \\ \midline
 Pitch shifting & 0.5 & transpose & -4 / 4 & semitones \\ \thinline
 Time stretching & 0.5 & transpose & 0.7 / 1.3 & ratio \\
 \bottomrule
\end{tabular}%
}
\caption{Training augmentation chains - Variant ($\mathcal{T}_V$) and Base ($\mathcal{T}_B$). We generate 4 augmentation for each anchor.}
\label{Tab: augmentations}
\end{table}

% \begin{table}[h]
% \centering
% \resizebox{1\linewidth}{!}{%
% \begin{tabular}{llll}
% \textbf{Augmentation} & \textbf{Probability} & \textbf{Parameter(s)} & \textbf{[Min,Max]} \\ \thickline
% \\
% \textbf{$\mathcal{T}_B$ (Base)} & & & \\ \midline
% Gain & 0.7 & Gain & [-15, 5] dB \\ \thinline
% Polarity inv. & 0.8 & - & - \\ \thinline
% Colored Noise & 0.8 & SNR, Spectral decay & [3, 30] dB, [-2, 2] dB/octave \\ \thinline
%  Low pass / High pass & 0.3 & Cutoff freq & [0.15, 7] kHz \\ 
%  Band pass / cut & 0.5 & Cutoff freq, Bandwidth & [0.15, 7] kHz, [0.5, 2] \\  \thinline
% Reverb & 0.5 & Room size & [0.2, 1] \\ \thinline
% Distortion & 0.6 & Drive & [1, 10] dB \\
% \\
% \textbf{$\mathcal{T}_V$ (Variant)} & & & \\ \midline
% Pitch shifting & 0.5 & Transpose & [-4, 4] semitones \\ \thinline
% Time stretching & 0.5 & Ratio & [0.7, 1.3] \\ \bottomrule
% \end{tabular}%
% }
% \caption{Training augmentation chains. Summary of Base ($\mathcal{T}_B$) and details of Variant ($\mathcal{T}_V$) transformations.}
% \label{Tab: augmentations}
% \end{table}

\subsection{MULE Architecture, LOEV++ Architecture}

\begin{figure}[h]
    \centering
    \includegraphics[width=1\linewidth]{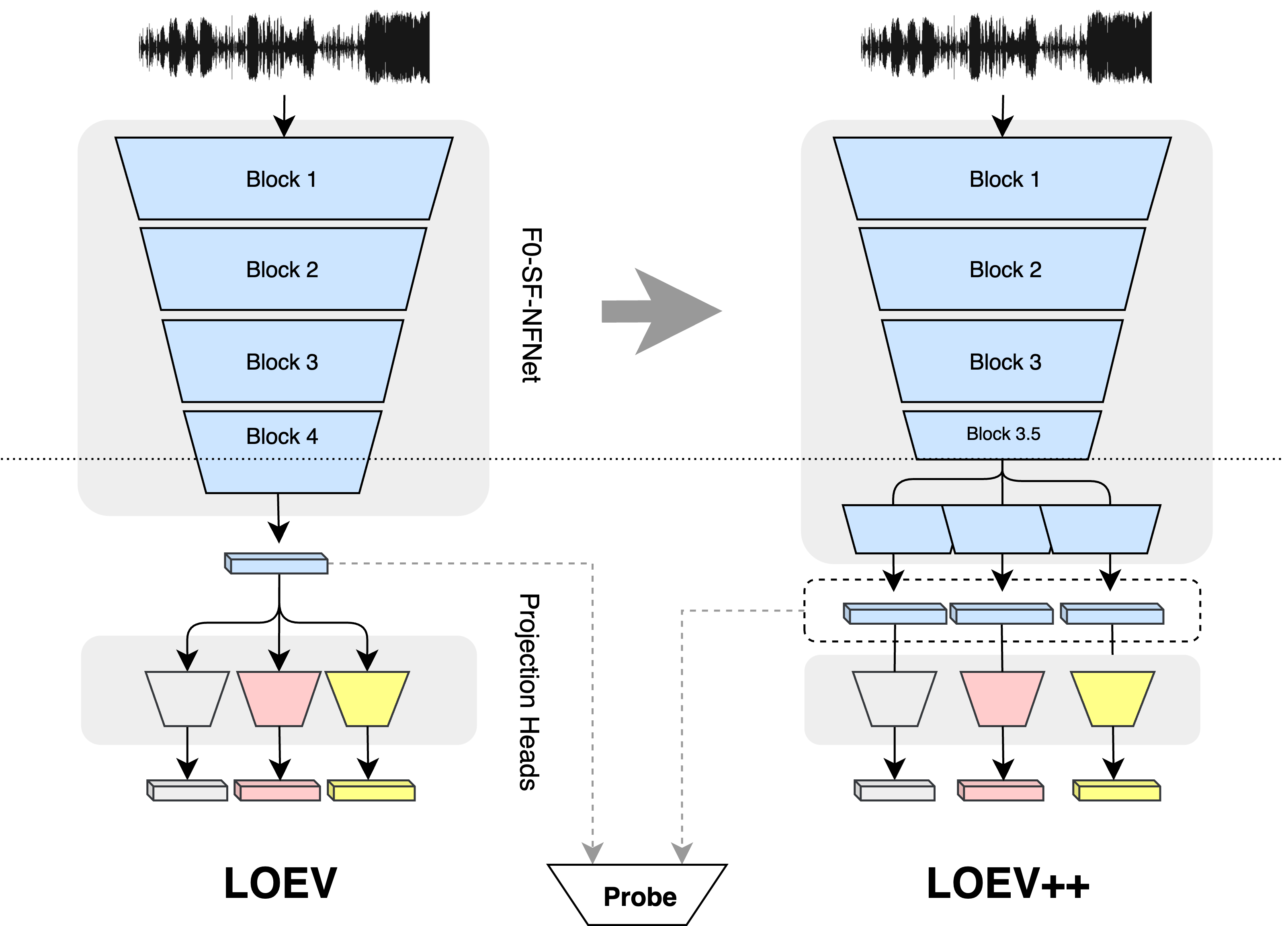}
    \caption{LOEV(++) architectures. In either case, the probing representation can come from the embedding superspace, or, for LOEV++, the concatenation of the parallel superspaces can be used, as in \cite{xiaoWhatShouldNot2021}}
    \label{fig:MuLOOC++}
\end{figure}

Tab. \ref{tab:architecture} Shows the architecture described in previous work \cite{wangLearningUniversalAudio2022, mccallumSupervisedUnsupervisedLearning2022}. 
We show the modified LOEV++ architecture in Fig. \ref{fig:MuLOOC++} - Half of the last block 4 of the SF-NF0Net architecture is parallelized into 3 blocks and prepended to the projection heads $g^i,g^p,g^t$. This changes our models' parameter count from 53M to 78M. Note that in LOOC \cite{xiaoWhatShouldNot2021}, parallelizing the \emph{conv5} block of the ResNet architecture close to \emph{triples the size of the network}. This might explain why the disentanglement results we obtain in Table \ref{tab:subspace} and Fig. \ref{fig:retrieval} can be perceived as modest. However, we wish to remain computationally sober when compared to MULE while achieving desirable disentanglement results. We could increase the portion of the network that is resorbed into the projection heads, but would lose that computational sobriety, and one might argue that training different contrastive models --- one per variant augmentation --- might be more sensible if we were to forgo all compute limitations.

\begin{table}[h]
\centering
\resizebox{1\columnwidth}{!}{
\begin{tabular}{cccc}
\hline
\textbf{Stage} & \textbf{Slow Path} & \textbf{Fast Path} & \textbf{$T \times F$} \\ \hline
Spectrogram & - & - & $400 \times 128$ \\ \hline
data layer & stride 4,1 & stride 1,1 & \begin{tabular}[c]{@{}c@{}}Slow : $100 \times 128$\\ Fast : $400 \times 128$\end{tabular} \\ \hline
Stem 1 & \begin{tabular}[c]{@{}c@{}}$1 \times 3$, 16\\ stride 1,1\end{tabular} & \begin{tabular}[c]{@{}c@{}}$3 \times 3$, 2\\ stride 2,2\end{tabular} & \begin{tabular}[c]{@{}c@{}}Slow : $50 \times 64$\\ Fast : $200\times 64$\end{tabular} \\
Stem 2 & \begin{tabular}[c]{@{}c@{}}$1 \times 3$, 32\\ stride 1,1\end{tabular} & \begin{tabular}[c]{@{}c@{}}$3 \times 3$, 4\\ stride 1,1\end{tabular} & \begin{tabular}[c]{@{}c@{}}Slow : $50 \times 64$\\ Fast : $200\times 64$\end{tabular} \\
Stem 3 & \begin{tabular}[c]{@{}c@{}}$1 \times 3$, 64\\ stride 1,1\end{tabular} & \begin{tabular}[c]{@{}c@{}}$3 \times 3$, 8\\ stride 1,1\end{tabular} & \begin{tabular}[c]{@{}c@{}}Slow : $50 \times 64$\\ Fast : $200\times 64$\end{tabular} \\
Stem 4 & \begin{tabular}[c]{@{}c@{}}$3 \times 3$, 128\\ stride 2,2\end{tabular} & \begin{tabular}[c]{@{}c@{}}$3 \times 3$, 16\\ stride 2,2\end{tabular} & \begin{tabular}[c]{@{}c@{}}Slow : $25 \times 32$\\ Fast : $100\times 32$\end{tabular} \\ \hline
Block 1 & $
\left[
\begin{array}{ccc}
1 \times 1, 128 \\
1 \times 1, 128 \\
1 \times 3, 128 \\
1 \times 1, 256 \\
\end{array}
\right] \times 1
$ & $
\left[
\begin{array}{ccc}
1 \times 1, 16 \\
1 \times 1, 16 \\
1 \times 3, 16 \\
1 \times 1, 32 \\
\end{array}
\right] \times 1
$ & \begin{tabular}[c]{@{}c@{}}Slow : $25 \times 32$\\ Fast : $100\times 32$\end{tabular} \\
Block 2 & $
\left[
\begin{array}{ccc}
1 \times 1, 256 \\
1 \times 1, 256 \\
1 \times 3, 256 \\
1 \times 1, 512 \\
\end{array}
\right] \times 2
$ & $
\left[
\begin{array}{ccc}
1 \times 1, 32 \\
1 \times 1, 32 \\
1 \times 3, 32 \\
1 \times 1, 64 \\
\end{array}
\right] \times 2
$ & \begin{tabular}[c]{@{}c@{}}Slow : $25 \times 16$\\ Fast : $100\times 16$\end{tabular} \\
Block 3 & $
\left[
\begin{array}{ccc}
1 \times 1, 768 \\
1 \times 1, 768 \\
1 \times 3, 768 \\
1 \times 1, 1536 \\
\end{array}
\right] \times 6
$ & $
\left[
\begin{array}{ccc}
1 \times 1, 96 \\
1 \times 1, 96 \\
1 \times 3, 96 \\
1 \times 1, 192 \\
\end{array}
\right] \times 6
$ & \begin{tabular}[c]{@{}c@{}}Slow : $25 \times 8$\\ Fast : $100\times 8$\end{tabular} \\
Block 4 & $
\left[
\begin{array}{ccc}
1 \times 1, 768 \\
1 \times 1, 768 \\
1 \times 3, 768 \\
1 \times 1, 1536 \\
\end{array}
\right] \times 3
$ & $
\left[
\begin{array}{ccc}
1 \times 1, 96 \\
1 \times 1, 96 \\
1 \times 3, 96 \\
1 \times 1, 192 \\
\end{array}
\right] \times 3
$ & \begin{tabular}[c]{@{}c@{}}Slow : $25 \times 4$\\ Fast : $100\times 4$\end{tabular} \\ \hline
\multicolumn{3}{c}{Global average pool \& concatenate} & $d_E = 1728$ \\ \hline
\end{tabular}}%
\caption{F0-SF-NFNet architecture as described in \cite{wangLearningUniversalAudio2022}.}
\label{tab:architecture}
\end{table}

\subsection{Training protocol discussion}
\label{appendix_subsection: training protocol discussion}

While our method is reasonably simple, it depends on a range of design choices which might impact its capability in learning robust representations and conserving information from all augmentations. The impact of some of these design choices will be explored, but some are too computationally expensive to make a comprehensive and exhaustive exploration of them feasible. A few straightforward examples are as follows:\\

\emph{Variant augmentation application probability} : Stronger augmentations are \emph{generally} beneficial to coarse music information retrieval tasks such as automatic tagging, as shown in previous work. In our case, for variant augmentations, as the probability of application gets higher, the contrastive matrix for a given head gets sparser, with certain application leading to an empty contrastive matrix, which is not desirable. There is certainly a balance to strike in terms of probability of application of each augmentation. \\

\emph{Similarity strategy}: Briefly mentioned in Section \ref{subsection: pretraining details}, the manner in which we consider augmented samples to be similar or not is an important design choice. If we simply take non-augmented pairs as positive, this does not cover same-parameter-augmented pairs as positives, which they are. Further, perceptually-similar augmented samples are treated the same as radically different augmented samples with our strategy, which might introduce confusion into the training objective.\\

\emph{Number of augmentations in $\mathcal{T}_V$}: As the number of variant augmentations gets higher, the model stores more relevant information about some of these augmentations, which might be beneficial for some downstream tasks. However, it seems unlikely that all possible augmentations can be accounted for, computationally, and from a model size standpoint. The more augmentations are included as variant augmentations, the more information the model must store about these augmentations, information which might be contradictory, and the necessary disentanglement limited by model size. Because we limit our study to two semantically-evident augmentations in the field of MIR, we do not consider this design choice to fall within the scope of this study.\\

\emph{In-track sampling strategy}. The position at which the positive pairs are taken within the track is also important in our case. MULE adopts an all-track sampling strategy where positive chunks can be sampled from anywhere within the track. In our case, we wish to preserve key and tempo information between two positives to avoid confusing the model with false positives. Indeed, key and tempo are often used to globally describe a track, but they are local properties that might change throughout a track. Tempo changes are relatively rarer, but key changes are fairly common. However, all-track sampling has been shown to be more robust for general music understanding \cite{choiProperContrastiveSelfsupervised2022}. To strike a balance, we design a sampling strategy where positives can either be sampled from \emph{the same} chunk, \emph{adjacent} chunks, or \emph{anywhere within the track}. By defauly these strategies are sampled uniformly but this is a debatable design choice, which might prove to deteriorate our variant head objectives.

\end{document}